\documentclass[runningheads]{llncs}

\usepackage{graphicx}
\usepackage{amsmath}
\usepackage{mathrsfs}
\usepackage[frozencache=true]{minted}
\usepackage{siunitx}
\usepackage{float}
\usepackage{bm}

\begin{document}
\title{Lettuce: PyTorch-based Lattice Boltzmann Framework}
%

\author{
Mario Christopher Bedrunka\inst{1,2} \and
Dominik Wilde\inst{1,2} \and
Martin Kliemank\inst{2} \and
Dirk Reith\inst{2,3} \and
Holger Foysi\inst{1} \and
Andreas Krämer\inst{4}\thanks{Corresponding author: kraemer.research@gmail.com}}
\authorrunning{Bedrunka et al.}
%
\institute{Department of Mechanical Engineering, University of Siegen, Paul-Bonatz-Straße 9-11, 57076 Siegen-Weidenau, Germany \and
Institute of Technology, Resource and Energy-efficient Engineering (TREE),\\ Bonn-Rhein-Sieg University of Applied Sciences, Grantham-Allee 20, 53757 Sankt Augustin, Germany\and
Fraunhofer Institute for Algorithms and Scientific Computing (SCAI), Schloss Birlinghoven, 53754 Sankt Augustin, Germany\and
Department of Mathematics and Computer Science, Freie Universität Berlin, Arnimallee 6, 14195 Berlin, Germany}
\maketitle              
\begin{abstract}
The lattice Boltzmann method (LBM) is an efficient simulation technique for computational fluid mechanics and beyond. It is based on a simple stream-and-collide algorithm on Cartesian grids, which is easily compatible with modern machine learning architectures. While it is becoming increasingly clear that deep learning can provide a decisive stimulus for classical simulation techniques, recent studies have not addressed possible connections between machine learning and LBM. Here, we introduce \emph{Lettuce}, a \emph{PyTorch}-based LBM code with a threefold aim. \emph{Lettuce} enables GPU accelerated calculations with minimal source code, facilitates rapid prototyping of LBM models, and enables integrating LBM simulations with \emph{PyTorch}'s deep learning and automatic differentiation facility. As a proof of concept for combining machine learning with the LBM, a neural collision model is developed, trained on a doubly periodic shear layer and then transferred to a different flow, a decaying turbulence. We also exemplify the added benefit of \emph{PyTorch}'s automatic differentiation framework in flow control and optimization. To this end, the spectrum of a forced isotropic turbulence is maintained without further constraining the velocity field. The source code is freely available from https://github.com/lettucecfd/lettuce.

\end{abstract}

\newpage

\section{Introduction}
\label{chapter:introduction}
Innovations are more important than ever in the face of global challenges such as climate change and pandemics. Bridging the gap from basic understanding to technological solutions often requires physical models in some stages of development. Such models can predict aspects of global warming or the spread of viruses based on fluid dynamics, for example \cite{porte2020,diwan2020,fabregat2021}. However, the models' equations, usually in the form of partial differential equations (PDEs), require efficient solution approaches due to their complexity. By using numerical methods, computers solve these PDEs, which are discretized in space and time. Even though great care is taken, numerical errors inevitably occur during discretization, where the truncation error usually depends on grid and time step size. Hence, large resolutions are necessary if high accuracy is required, especially if modeling is reduced as much as possible, e.g., during direct numerical simulations (DNS). When DNS are computationally intractable, alternatives like the well-known Reynolds averaged Navier-Stokes equations (RANS) or large-eddy simulation (LES) are routinely used. These approaches require expressing unresolved or unclosed terms through known quantities \cite{Wilcox,Sagaultbook}. Parameters involved in these modeling approaches are often not optimally determined or change for different flows. For this purpose, machine learning (ML) in fluid dynamics is a rapidly evolving research field that enables new impulses to tackle such problems and provide interesting new approaches to the modeling problem. The benefit of ML approaches in computational fluid dynamics (CFD) simulations has been demonstrated in various recent studies  \cite{FONT2021110199,Brunton2020,kochkov2021,Um2020}. A successful use of machine learning in combination with well-known Navier-Stokes solvers was recently shown by Kochkov et al. \cite{kochkov2021}. They introduced learned forcing terms into PDE solvers, thereby reaching the same accuracy as a classical solver at 8-10x finer resolution. \\[-.35cm]

When integrating machine learning into numerical algorithms, it can be beneficial to resort to simulation methods whose mathematical structure is easily compatible with neural networks. The present work shows that a highly suitable CFD approach for this purpose is the lattice Boltzmann method \cite{McNamara1988,Kruger2016} (LBM), which is particularly competitive in the fields of transient, turbulent, or multiphase fluid dynamics. The LBM is a second order accurate simulation method that exhibits similar performance as classical finite difference schemes \cite{wichmann2021}. In contrast to classical solvers, it involves a linear streaming of particle distribution functions on a regular grid and a local collision step. Despite its successful application to many fluid problems, recent studies have only scarcely addressed possible combinations of ML and LBM. As a prototypical approach, Hennigh \cite{hennigh2017} has demonstrated the potential of ML-driven flow prediction based on LBM. He compressed flow fields onto coarser grids using convolutional autoencoders and learned the propagation of the latent representations, which can then be decoded back onto the fine grid. This approach, however, has limited transferability as it is primarily data-informed and does not encode the underlying physics. Furthermore, Rüttgers et al. \cite{ruttgers2020} have applied deep learning methods to the lattice Boltzmann method to predict the sound pressure level caused by objects. They introduced an encoder-decoder convolutional neural network and discussed various learning parameters to accurately forecast the acoustic fields. To the best of our knowledge, no further ML-enhanced LBM methods were proposed. \\

Although the mathematics and physics behind the LBM are ambitious, the implementation is relatively simple. This allows beginners to quickly set up simple test cases in one or two dimensions with popular scripting languages like Matlab or Python. However, when switching to three dimensions, the algorithmic and computational complexity of simulations severely limits such prototypical implementations. More efficient simulations in compiled languages usually rely on optimized software packages and require initial training to simulate complex flows. This is particularly true for GPU-accelerated codes, which enhance performance \cite{Obrecht2013,Lenz2019} but defy fast implementation. \\[-.3cm]

Both the lack of machine learning studies in the context of LBM and the code complexity of 3D implementations motivate an LBM framework that allows ease of use, despite extensive built-in functionality for machine learning algorithms. For this purpose, the software package \emph{Lettuce} has been developed based on the open-source machine learning framework \emph{PyTorch} \cite{pytorch}. \emph{PyTorch} implements optimized numerical operations on CPUs and GPUs, which can easily be accessed via Python instructions. Internally, those operations are vectorized using efficient backends such as BLAS/LAPACK and highly optimized CUDA code.
By resorting to those efficient \emph{PyTorch} core routines, the LBM code remains maintainable and lean, which will be demonstrated throughout this article. Furthermore, the \emph{Lettuce} framework can seamlessly integrate \emph{PyTorch}'s machine learning modules into the fluid dynamics solver and thereby provide a substantial contribution to future work in this area. \\[-.3cm]

\emph{Lettuce} is meant to complement existing optimized codes
like the well-known LBM frameworks OpenLB \cite{heuveline2010openlb}, Palabos \cite{Latt2020}, waLBerla  \cite{Bauer2020,Godenschwager2013}, and others \cite{Schmieschek2017,Mora2020,Pastewka2019}.
It intends to bridge the gap between scripting language codes for local machines and highly optimized codes based on compiled programming languages that run efficiently on computer clusters, too. With an off-the-shelf GPU, \emph{Lettuce} can simulate fairly complex three-dimensional problems even on a local workstation with 24 GB of GPU memory. Independent of machine learning research, this also enables rapid prototyping of general methodological extensions for the lattice Boltzmann method. \\
This short paper is structured as follows. Section \ref{chapter:description} presents an overview of the software functionalities and briefly describes the LBM and its implementation. Section \ref{chapter:Advanced Functionalities} shows simple examples of coupling the LBM with machine learning, demonstrates the use of \emph{PyTorch}'s automatic differentiation capabilities in CFD simulations, and provides a computational benchmark. Section \ref{chapter:conclusion} presents a summary and conclusions. Scripts for all simulations in this paper are accessible on https://github.com/lettucecfd/lettuce-paper.

\section{Software Description}
\vspace{-.1cm}
\label{chapter:description}
\subsection{Software Functionalities}
\label{chapter:functionalities}
The lattice Boltzmann method (LBM) is based on the kinetic theory of gases, concretely a discretized version of the BGK-Boltzmann equation. 
It evolves a discrete particle distribution function $f_{i}\left(\mathbf{x},t\right)$ according to the lattice Boltzmann equation
\begin{equation}
    f_{i}\left(\mathbf{x}+\mathbf{c}_{i}\delta_t, t+\delta_t \right) = f_{i}\left(\mathbf{x},t \right) + \Omega_{i}\left(\bm{f}(\mathbf{x}, t) \right),
\end{equation}
where $\delta_t$ and $\mathbf{c}_i$ denote the time step and discrete particle velocities, respectively.
At its core, the LBM involves two operations. First, the so-called streaming step shifts the particle distributions $f_{i}\left(\mathbf{x},t\right)$ to the neighboring nodes along the trajectories $\mathbf{c}_i$. Second, the collision step introduces interactions between the particles on each node. Among the various collision models available in the literature, $\Omega\left(\bm{f}\right)$ is selected based on considerations such as asymptotic behavior, accuracy, memory usage and stability. The most commonly used collision operator is the Bhatnagar-Gross-Krook model:\vspace{-.2cm}
\begin{equation}
    \Omega_{i}\left(\bm{f}\right) = -\frac{f_{i}-f_{i}^{\text{eq}}}{\tau}
\end{equation}
This operator describes the relaxation of the particle distribution function towards an equilibrium distribution influenced by a single relaxation parameter $\tau$. The equilibrium distribution is given by
\begin{equation}
    f_{i}^{\text{eq}}\left(\mathbf{x},t\right) = w_{i}\rho\left( 1 + \frac{\mathbf{u \cdot c}_{i}}{c_{s}^{2}} + \frac{\left(\mathbf{u \cdot c}_{i}\right)^{2}}{2c_{s}^{4}} - \frac{\mathbf{u}\cdot\mathbf{u}}{2c_{s}^{2}} \right),
\end{equation}
where $w_{i}$ and $c_s$ are the lattice weights and speed of sound, respectively. The density $\rho$ and fluid velocity $\mathbf{u}$ are obtained as 
\begin{equation}
    \rho \left( \mathbf{x}, t\right) = \sum_{i} f_{i} \left(\mathbf{x},t \right) 
    \qquad \mathrm{and} \qquad 
    \rho \mathbf{u}\left( \mathbf{x}, t\right) = \sum_{i} \mathbf{c}_{i}f_{i} \left(\mathbf{x},t \right) .
    \vspace{-.3cm}
\end{equation}
\emph{Lettuce} is equipped with a variety of frequently used collision models, such as  the Bhatnagar-Grook-Krook (BGK) model \cite{Bhatnagar1954}, multi-relaxation time collision models \cite{Lallemand.2000,Dellar.2003}, the two-relaxation time model \cite{Ginzburg2008}, the regularized model \cite{Latt2006} and entropic two-relaxation time models by Karlin, Bösch and Chikatamarla (KBC) \cite{Karlin2014}. For the latter, implementations are rare in open software packages. Many of these collision models are implemented in a stencil- and dimension-independent manner.

\begin{figure}[b!]
    \centering
    \includegraphics[width=0.28\linewidth]{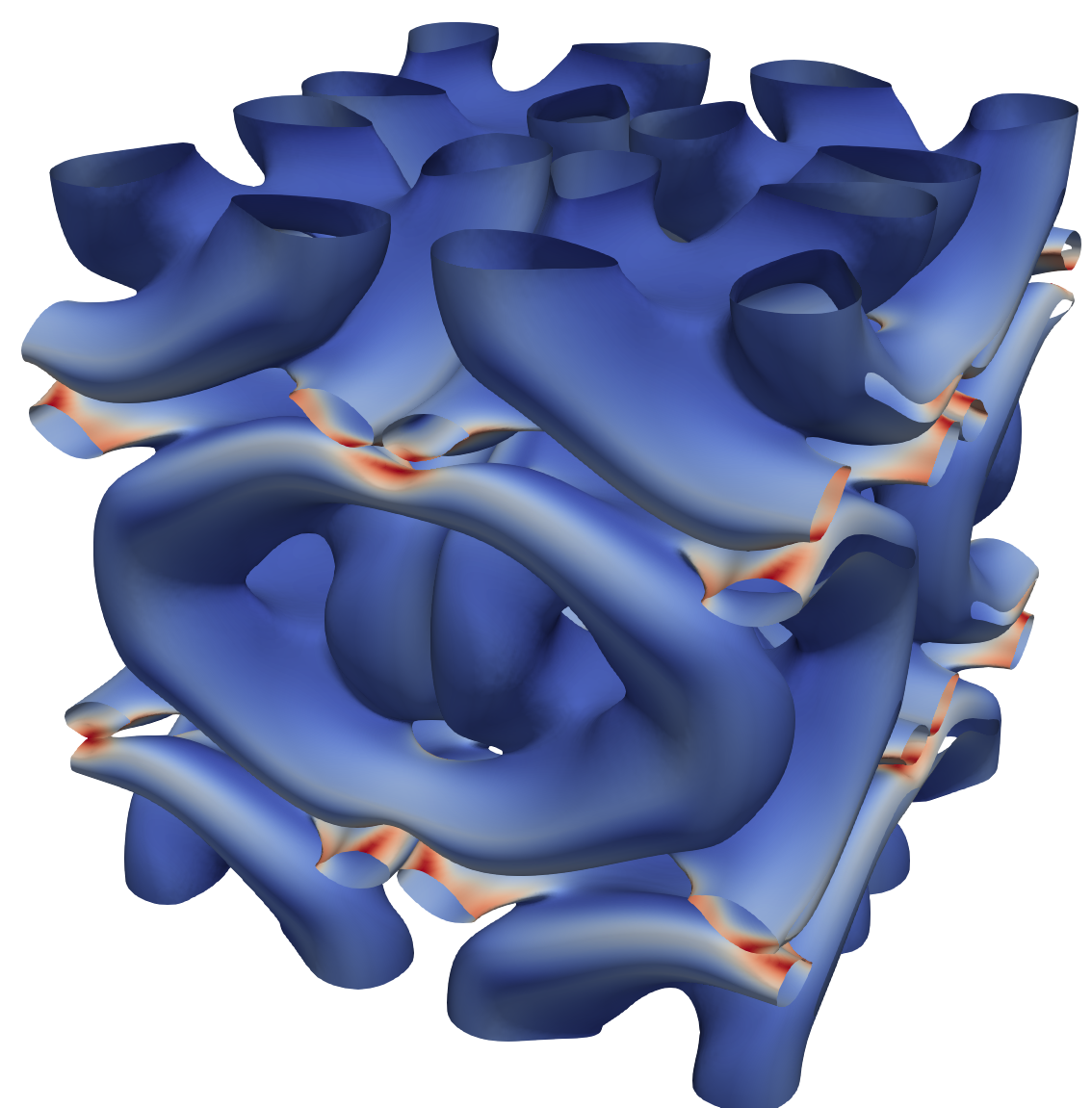} \hspace{0.06\linewidth}
    \includegraphics[width=0.28\linewidth]{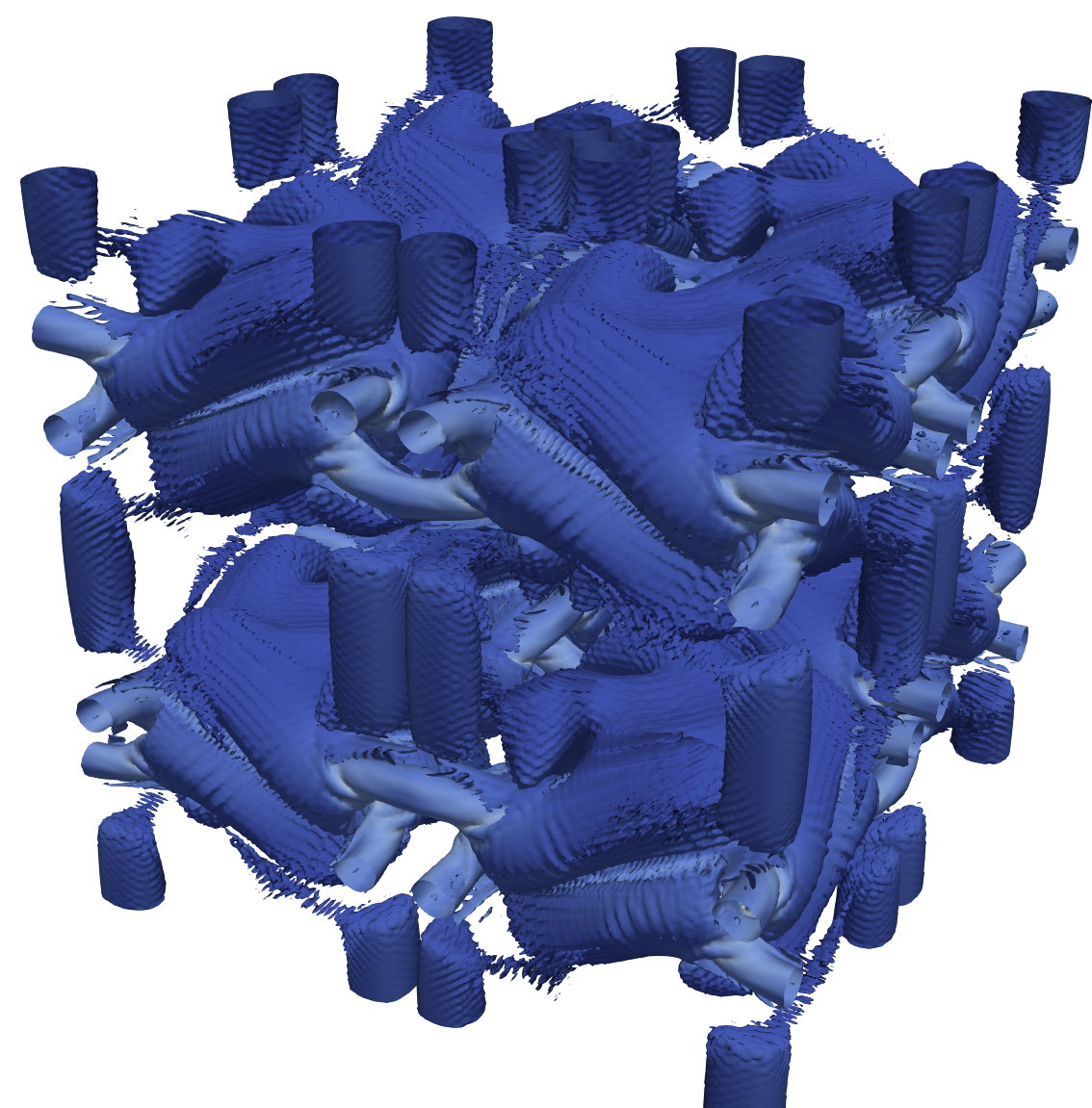} \hspace{0.06\linewidth}
    \includegraphics[width=0.28\linewidth]{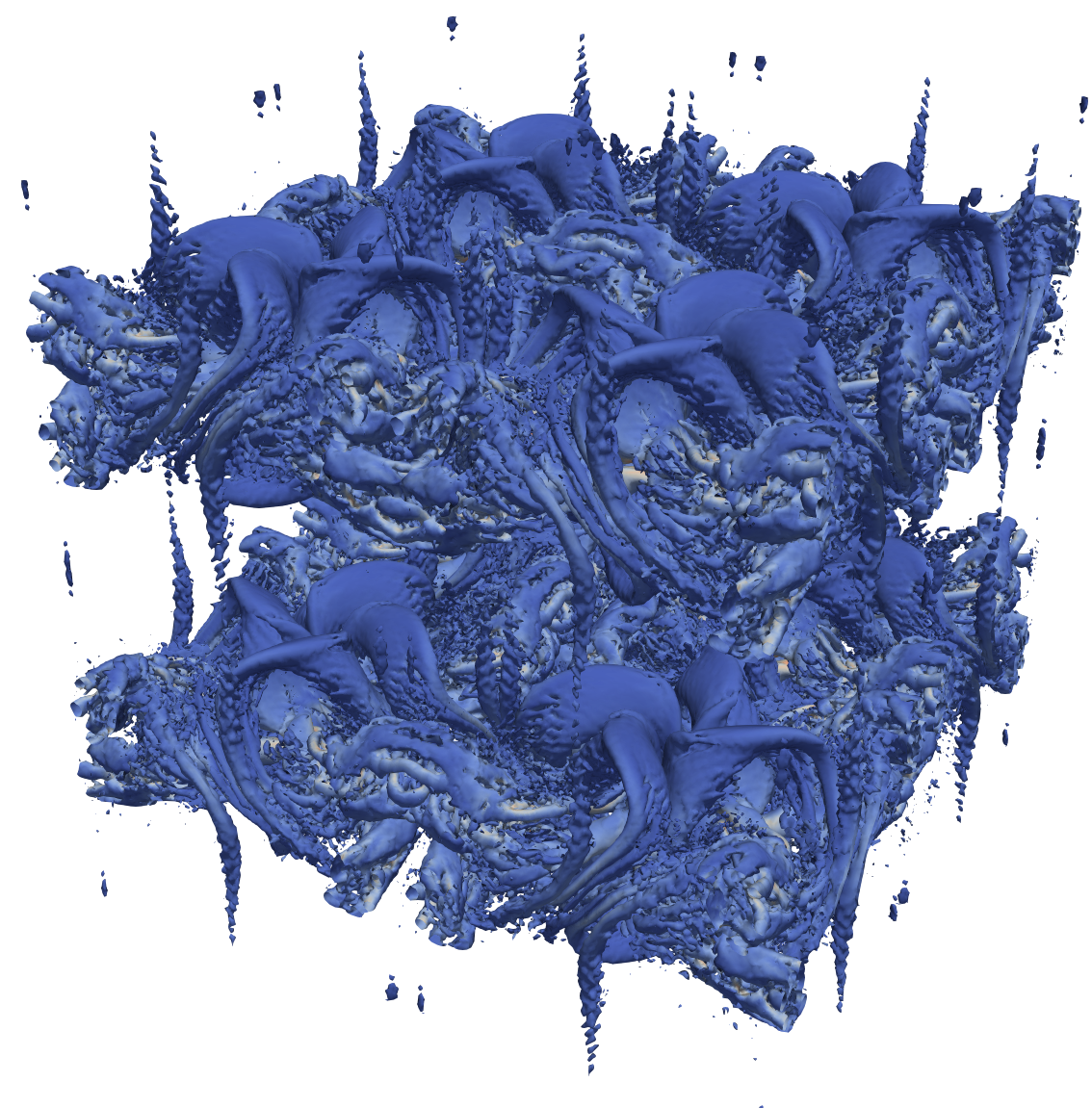}
    \caption{Q-criterion isosurfaces of a three-dimensional Taylor-Green Vortex at time step t=5000, 7000 and 10000 colored by streamwise velocity. The Reynolds number and grid resolution were $1600$ and $256^3$, respectively.}
    \label{fig:tgv3d_Q-criterion}
\end{figure}

\subsection{Code example}
\label{chapter:example}
After \emph{Lettuce} is installed, the user can run simulations with minimal 
code. The following example demonstrates a lean executable Python script that simulates a three-dimensional Taylor-Green vortex (TGV3D), one of several flows provided in the library. The \emph{Lettuce} library contains various boundary conditions, forcing and initialization schemes, thus covering a wide range of setups. After importing \emph{Lettuce}, the stencil and the hardware are selected. Then, the flow, collision model, and streaming step are chosen and run through the \texttt{Simulation} class.

\begin{minted}{Python}
import lettuce as lt
lattice = lt.Lattice(stencil=lt.D3Q27, device='cuda') #or 'cpu'
flow = lt.TaylorGreenVortex3D(
    resolution=256,
    reynolds_number=1600,
    mach_number=0.05,
    lattice=lattice)
collision = lt.BGKCollision(
    lattice=lattice,
    tau=flow.units.relaxation_parameter_lu)
streaming = lt.StandardStreaming(lattice)
simulation = lt.Simulation(
    flow=flow,
    lattice=lattice,
    collision=collision,
    streaming=streaming)
simulation.step(num_steps=10000)
\end{minted}

\emph{Lettuce} provides various observables that can be reported during the simulation (e.g. kinetic energy, enstrophy, energy spectrum). These observables can be added easily to the \texttt{Simulation} class and exported for further analysis as follows: 

\begin{minted}{Python}
energy = lt.IncompressibleKineticEnergy(lattice, flow)
simulation.reporters.append(
    lt.ObservableReporter(energy, interval=10,))
simulation.reporters.append(
    lt.VTKReporter(
        lattice, 
        flow, 
        interval=10, 
        filename_base="./output"))
\end{minted}

\noindent Besides, \emph{Lettuce} comes with a VTK-reporter based on the PyEVTK library \cite{herrera}. This reporter exports velocity components and pressure, which both can then be visualized by third-party software. An example is given in Figure \ref{fig:tgv3d_Q-criterion}, which shows the isosurfaces of the three-dimensional Taylor-Green vortex simulation from the code snippet above. 

\begin{figure}[ht]
    \centering
    \includegraphics[width=0.49\linewidth]{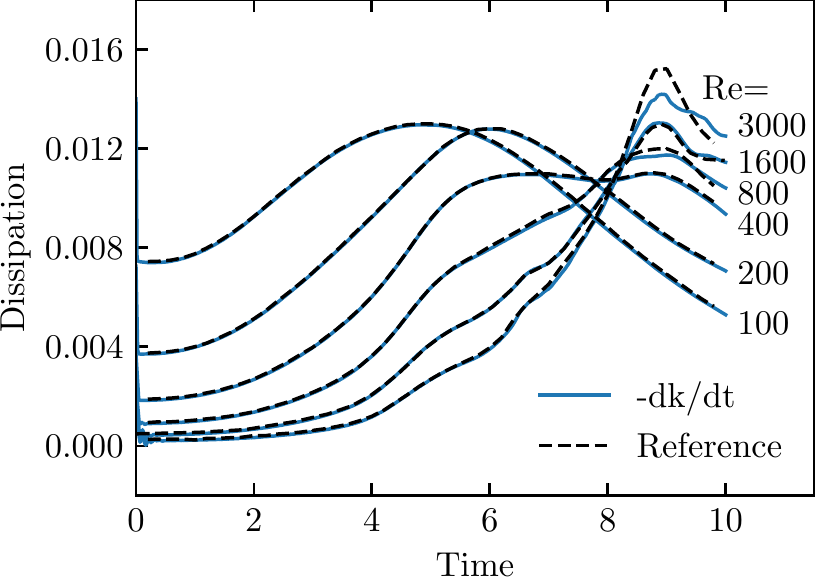}
    \vspace{-.3cm}
    \caption{Energy dissipation rate $\epsilon (t)=-dk/dt$ of a three-dimensional Taylor-Green-Vortex using the BGK collision model. Reference is taken from Brachet \cite{brachet1983}.}
    \label{fig:tgv3d_dissipation}
\end{figure}

Figure \ref{fig:tgv3d_dissipation} shows the energy dissipation that is obtained from the kinetic energy $k$ by calculating $-dk/dt$ (=$\nu \langle\epsilon\rangle$ in isotropic turbulence) using finite differences. That way it includes numerical dissipation effects and  spurious contributions from LBM, too. The data is compared to the reference taken from Brachet \cite{brachet1983}, who used a spectral code with a resolution of $256^{3}$ grid points. The dissipation, $-dk/dt$, shows excellent agreement with the reference data up to a Reynolds number of 1600. For Reynolds numbers of $Re = 3000$ and higher the maximum dissipation rate deviates slightly due to under-resolution.\\
\section{Advanced Functionalities}
\label{chapter:Advanced Functionalities}
\subsection{Machine Learning}
\label{chapter:deeplearning}
The collision model constitutes the core of the LBM as it determines the macroscopic system of PDEs and thereby encodes the solver's physics. It has long been known that the choice of collision model for a given PDE is ambiguous, which is related to the discrepancy between the number of degrees of freedom (discrete distribution functions) and macroscopic variables of interest. For example, the standard D2Q9 stencil uses nine degrees of freedom per lattice node to encode six physically relevant moments (density, momentum, and stress tensor). The remaining degrees of freedom (represented by higher-order moments or cumulants) are usually propagated in a way that offers certain numerical advantages such as improved stability \cite{Karlin2014,Latt2006,kramer2019} or accuracy \cite{geier2017}.\\ 

In the following, we want to exploit this ambiguity to define neural collision operators as a more accurate alternative to classical collision models. For this purpose, we define a collision model that relaxes the moments $m_i$ towards their respective equilibria $m^{\text{eq}}_{i}$ by individual relaxation rates $\mathbf{S}=\text{diag}\left( \tau_0, \tau_1, ...,\tau_{Q-1}\right)$, where $Q$ is the number of discrete velocities per grid point. A transformation matrix $\mathbf{M}$ (according to Dellar \cite{Dellar.2003}) maps the distribution function $\bm{f}$ to the moments $\mathbf{m}=\mathbf{M}\bm{f}=(\rho, \rho\mathbf{u},\mathbf{\Pi}, \mathscr{N}, \pmb{\mathscr{J}})^T$, where $\rho$ is the density, $\mathbf{u}=(u,v)$ is the fluid velocity, $\mathbf{\Pi}$ is the momentum flux, and $\mathscr{N}$ and $\pmb{\mathscr{J}}=(\mathscr{J}_{0},\mathscr{J}_{1})$ are non-hydrodynamic moments. These moments are relaxed towards the moment equilibrium $\mathbf{m}^{\mathrm{eq}}=\mathbf{M}\bm{f}^\mathrm{eq}$ with the relaxation rates given by $\mathbf{S}$:
\begin{equation}
    \Omega(\bm{f}) = -\bm{M}^{-1}\bm{S}^{-1}(\bm{M}\bm{f} - \bm{m}^\mathrm{eq}).
\end{equation}
The relaxation rates for the conserved moments $\rho$ and $\rho\bm{u}$ have no effect and are thus set to unity. Since the shear relaxation rates are related to the kinematic viscosity $\nu$, they are set to $\tau_n = \nu c_{s}^{-2}\delta_t^{-1}+0.5,$ $n=3,4,5,$ which recovers the Navier-Stokes equations in the weakly compressible regime. A neural network provides the relaxation rates $\tau_{n}$, $n=6,7,8,$ for the higher moments. For this purpose, a shallow network with one hidden layer and 530 parameters is introduced to keep the computational cost feasible. The network determines the higher-order relaxation rates based on local moments and is optimized to reproduce a finer-resolved reference simulation. Moreover, we want to ensure that the collision operator is stable. For this purpose, an exponential function is applied to the output $\tilde{\tau}_{n}$ of the neural network:  $\mathbf{\tau}_{n}=\mathrm{exp}(\tilde{\tau}_{n})+0.5$, $n=6,7,8.$ This operation renders the output larger than 0.5, which prevents excessive over-relaxation. The relaxation parameters are not upper bounded as $\tau_n\to \infty$ yields the identity and is usually uncritical in terms of stability. \\[-.3cm]

Training data was generated by simulating a doubly periodic shear layer at a Reynolds number of 5000 on a domain of size $128^2$ using the BGK model \cite{brown1995}. The shear layer provides both large gradients and smooth areas which need to be detected. Most relevant engineering flows have features that are locally present in shear layer turbulence, such that good performance of a model in this setup will likely transfer to other flows. Instead, training with isotropic flows only would likely hamper transferability.

Depending on the information contained in the local moments, the network should adjust the local relaxation parameters. The training procedure optimizes the network parameters $\bm\vartheta$ to minimize the discrepancy between a low-resolution and a high-resolution simulation. Therefore, the discrete distributions from the training trajectory are mapped onto a coarser grid with $64^2$ grid points. Batches of short simulations with the neural collision model are started from each snapshot. After 100 simulation steps, the flow fields are compared with the finer-resolved reference simulation based on energy $E$, vorticity $\omega$, and velocity $u,$ which are all computed on the coarse grid. The mean-squared error (MSE) of these quantities from the reference are minimized using the Adam optimizer and the loss function\vspace{-.3cm}
\begin{equation}
    \begin{split}
    L\left(t; \bm{\vartheta} \right) =\quad & w_u \sum_{i}^{N} \text{MSE} \left( u(\mathbf{x}_i,t;\bm{\vartheta}),\tilde{u}(\mathbf{x}_i,t)\right) \\
    + &w_\omega \sum_{i}^{N} \text{MSE} \left( \omega(\mathbf{x}_i,t;\bm{\vartheta}),\tilde{\omega}(\mathbf{x}_i,t)\right) \\
    + &w_E \text{ MSE} \left( E(t;\bm{\vartheta}),\tilde{E}(t)\right),
    \end{split}
    \label{eq:loss}
\end{equation}
where $N$ is the number of grid points. The weights are hyperparameters that were selected as $w_u:=0.6,$ $w_\omega := w_E := 0.2.$ This choice emphasizes the optimization of dissipation effects, which are critical in under-resolved turbulence. Such flows exhibit large gradients that occur intermittently, leading to locally under-resolved spatial structures. Therefore, the model has to strike a balance between retaining the physical structures on small scales and adding numerical dissipation for stabilization.

The fluid velocity is the most natural target as it directly measures numerical errors. The kinetic energy tracks the dissipation globally but does not resolve the spatial heterogeneity. In contrast, including the vorticity as a target stresses the finest resolved structures; the enstrophy, i.e., the integral over the vorticity magnitude, measures the viscous shear stresses that induce dissipation. In homogeneous turbulence, it peaks at high wave numbers around the Kolmogorov scale.  Consequently, optimizing the loss function \eqref{eq:loss} deliberately targets the dissipation occurring on small scales. A detailed hyperparameter optimization, the inclusion of other target properties, and the incorporation of multiple flows with different boundary conditions in the training set will likely further improve the model. These next steps as well as a systematic study of transferability are beyond the scope of this proof-of-concept and will be left for future work.

\begin{figure}[htbp]
    \centering
    \includegraphics[width=0.49\linewidth]{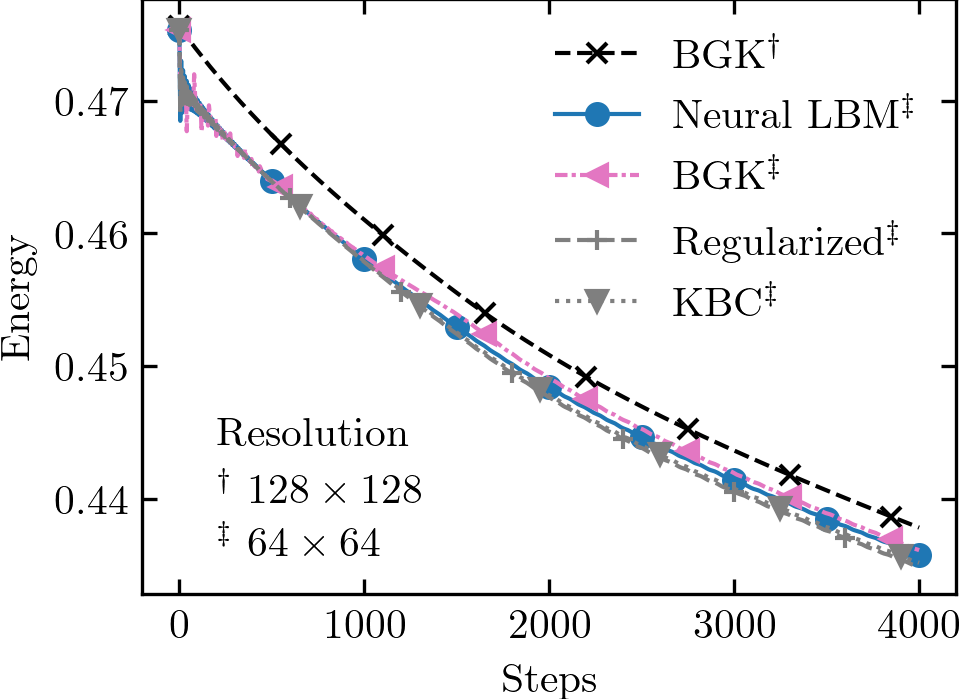}
    \caption{Comparison of the kinetic energy evolution of various collision models for the doubly periodic shear layer at $Re=5000$.}
    \label{fig:doublyshearlayer_stats}
\end{figure}
\begin{figure}[htbp]
    \centering
    \includegraphics[width=0.79\linewidth]{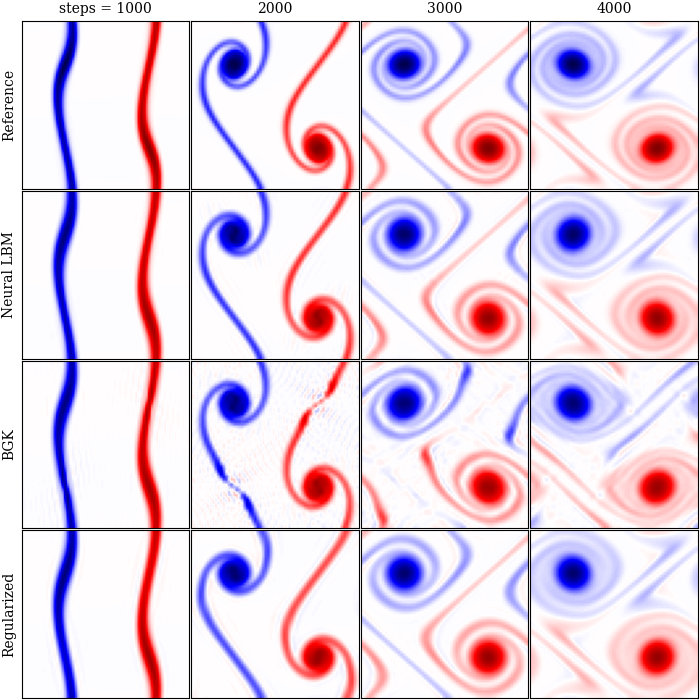}
    \caption{Evolution of vorticity fields for a doubly periodic shear layer flow for Reynolds number 5000 using several collision models \cite{brown1995}.
    }
    \label{fig:evolution_doublyshearlayer}
\end{figure}

\begin{figure}[hbtp]
    \centering
    \includegraphics[width=0.99\linewidth]{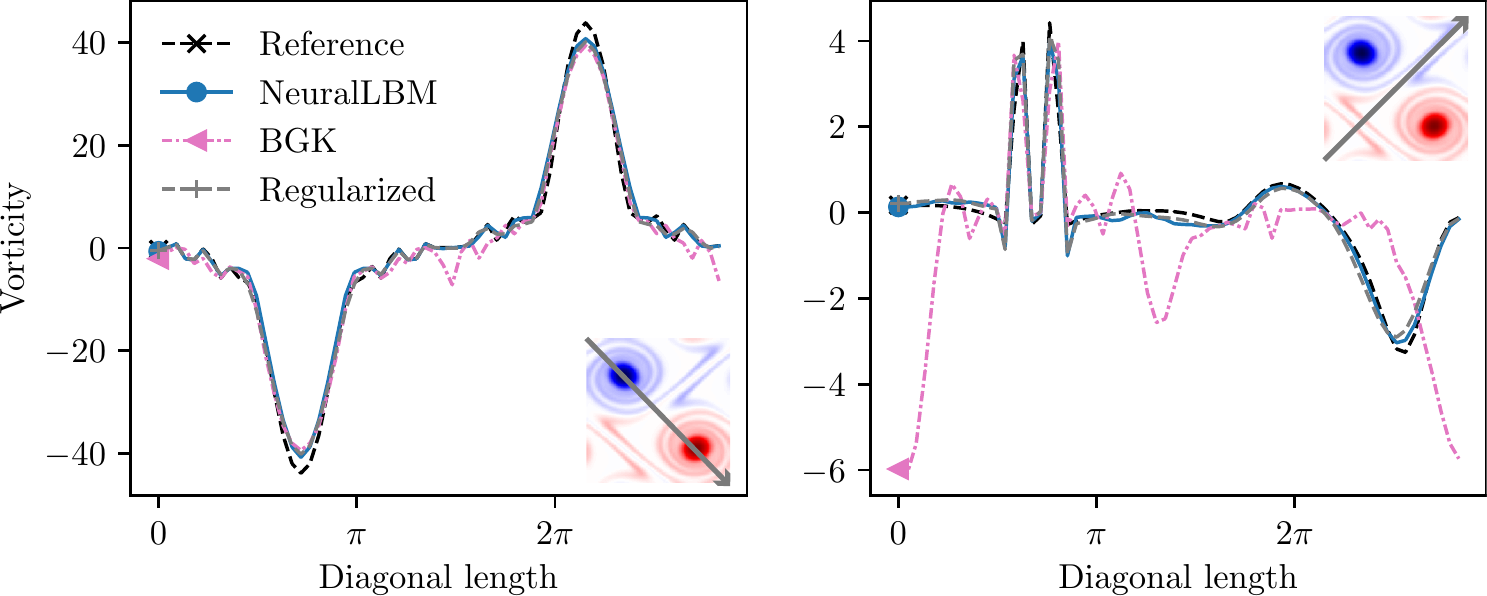}
    \caption{Vorticity along a diagonal line for a doubly periodic shear layer flow after 4000 steps for Reynolds number 5000.}
    \label{fig:vorticity_diag}
\end{figure}

Turning to results, Figure \ref{fig:doublyshearlayer_stats} compares the evolution of kinetic energy for the doubly periodic shear layer. Several collision models were used for this assessment. In the beginning, the energy of the lower-resolved simulations dropped due to under-resolution. Then, all collision models produced similar energies. However, the vorticity fields depicted in Figure \ref{fig:evolution_doublyshearlayer} clearly show that the simulation using the BGK operator can no longer capture the vortices from the finer-resolved reference simulation. In contrast, the neural collision model accurately reproduces these structures, as shown in more detail in Figure \ref{fig:vorticity_diag} by the vorticity along the diagonals. The improvement compared to the BGK operator becomes clear while still providing less dissipation than the regularized model.\\[-.3cm]

\begin{figure}[htbp]
    \centering
    \includegraphics[width=0.46\linewidth]{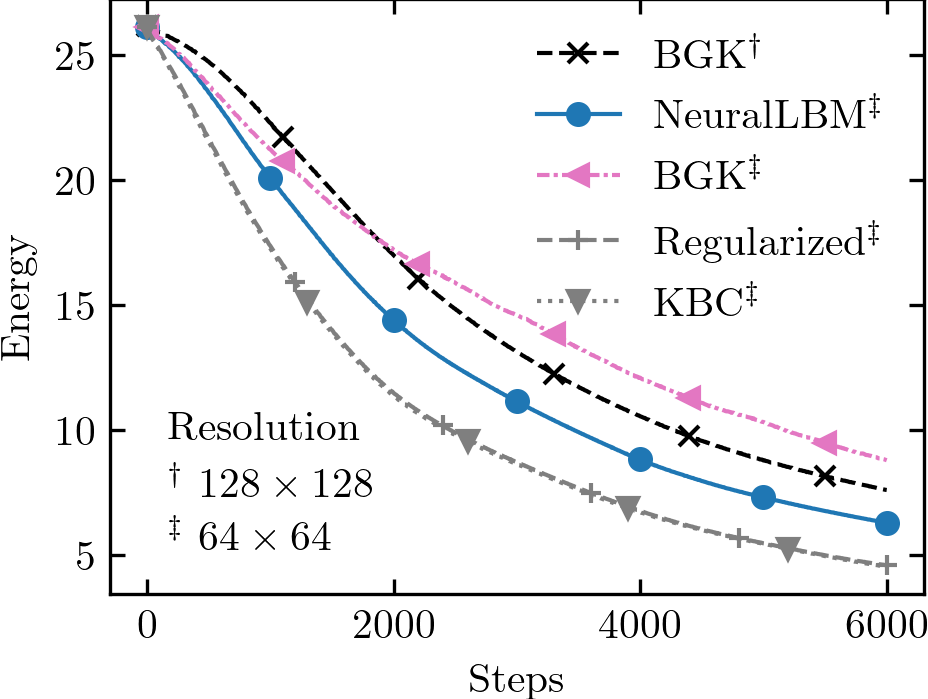}
    \includegraphics[width=0.49\linewidth]{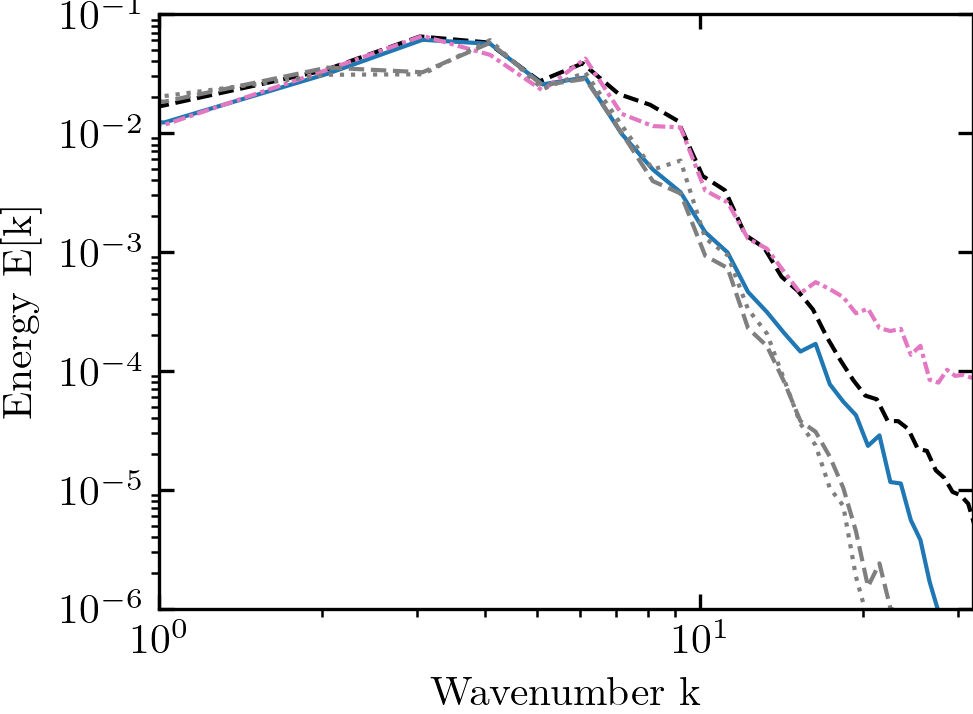}
    \caption{Enstrophy evolution (\textit{left}) and energy spectrum (\textit{right}) of an isotropic decaying turbulence at Reynolds number 30000. Colors and line types are equal for both plots.\vspace{-.7cm}}
    \label{fig:decaying_stats_re30000}
\end{figure}

\begin{figure}[htb]
    \centering
    \includegraphics[width=0.79\linewidth]{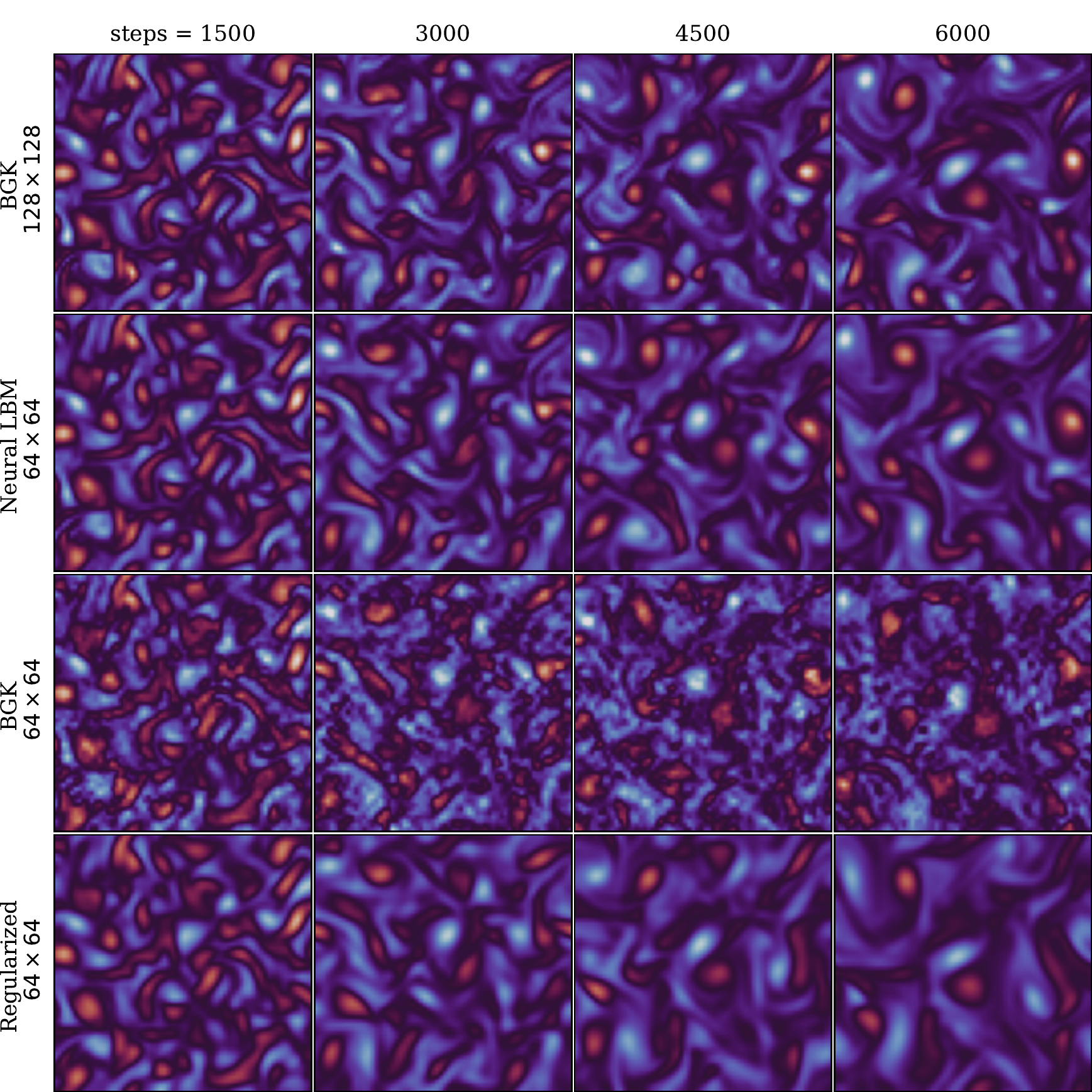}
    \caption{Evolution of vorticity fields for a isotropic decaying turbulence flow for Reynolds number 30000 using several collision models.}
    \label{fig:evolution_decaying_re30000}
\end{figure}

The crucial question is whether the optimized network is transferable to other flows. Figure \ref{fig:decaying_stats_re30000} shows the vorticity evolution and energy spectrum for an isotropic decaying turbulence simulation at a Reynolds number of 30000 \cite{samtaney2001}. Although trained on a different flow, the neural collision model reproduced the vortex field far better, while other collision models were either unstable or overly dissipative, as shown in Figure \ref{fig:evolution_decaying_re30000}. The BGK model was not able to handle the high Reynolds number and introduced unphysical small-scale oscillations. These numerical artefacts are visible in the energy spectrum, revealing a lot of unphysical energy accumulated at high wavenumbers. By contrast, the KBC and regularized collision models are more dissipative at larger wavenumbers, resulting in much faster energy and enstrophy decay. In comparison to these baseline models, the ML-enhanced simulation produced the best match with the reference simulation. This example demonstrates generalization capabilities and the potential benefit of using collision models based on neural networks.

A promising future direction of research is to target the current limitations of the LBM, including high Mach number compressible flows. These flows require higher-order moments so that current compressible LBMs usually resort to larger stencils \cite{Coreixas2020,Frapolli2015,Latt2020a,Wilde2021}, off-lattice streaming \cite{Chen2020,Wilde2021,Wilde2020}, or non-local collision models \cite{Saadat2021,Saadat2019} that introduce additional numerical approximations. In this application, neural collisions could help reduce numerical errors and advance the state of the art.

\subsection{Flow Control through Automatic Differentiation}
\label{chapter:automaticdiff}
The availability of an automatic differentiation framework within a CFD simulation engine has additional advantages (besides machine learning). \emph{PyTorch} provides analytic derivatives for all numerical operations, which is, for example, useful in flow control and optimization. \\
As a demonstration of these capabilities, forced isotropic turbulence is simulated, i.e., the energy spectrum is maintained throughout the simulation using a forcing term as detailed below. A cost functional $R$ is introduced as the relative deviation of the instantaneous spectrum $\sigma(\mathbf{u})$ from the target spectrum $\sigma_0$ with a cutoff at $c:=2\cdot 10^{-5}.$ $R$ is defined as
$$
R(\bm{u}) = \| \ln\max(\sigma(\bm{u}), c) - \ln\max(\sigma_0, c) \|_2^2,
$$
where the logarithm and maximum are taken elementwise. \\[-.3cm]

To incorporate this restraint into the simulation, the equilibrium distribution $f^\mathrm{eq}(\rho, \mathbf{u}+\Delta \mathbf{u})$ is expanded around a velocity that is locally shifted by a forcing term $\Delta \mathbf{u} := -\kappa \cdot \nabla_\mathbf{u} R$ with a force constant $\kappa=5\cdot 10^{-5}.$ 
Computing the gradient requires differentiating through various numeric operations, including a Fast Fourier Transform, which is easily done within \emph{PyTorch} due to the automatic differentiation facility.

\begin{figure}[htbp]
    \centering
     \begin{minipage}[t]{0.34\textwidth}
     \includegraphics[width=1\linewidth]{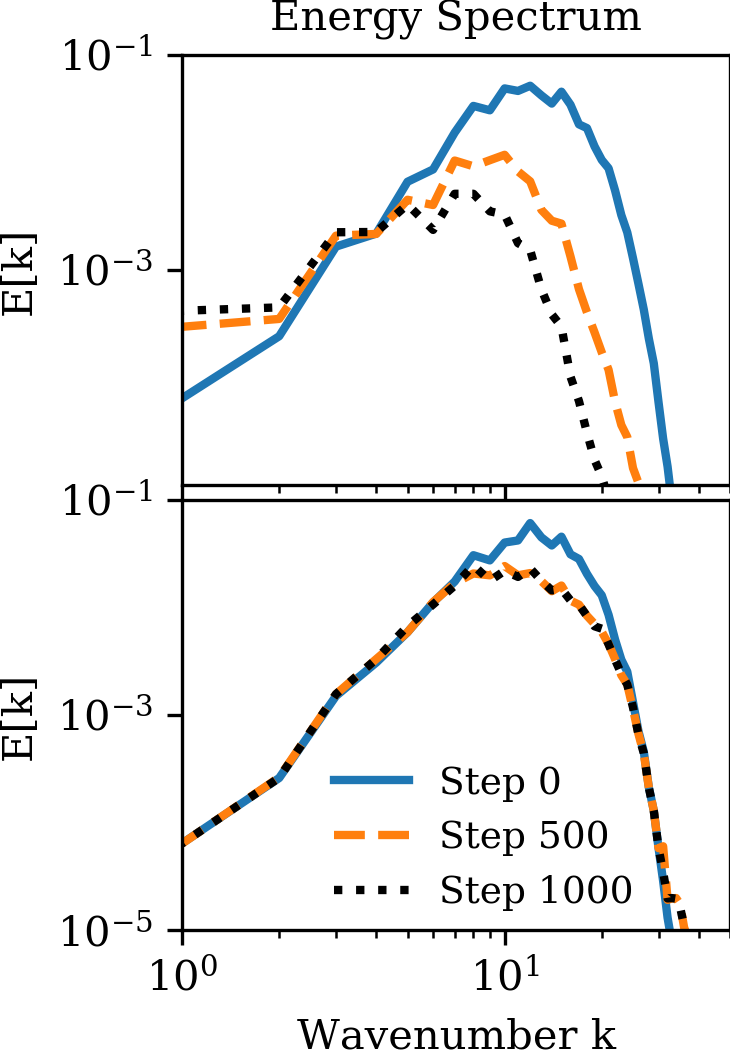}
     \end{minipage}\hfill
     \begin{minipage}[t]{0.65\textwidth}
     \vspace{-2.36in}
     \includegraphics[width=1\linewidth]{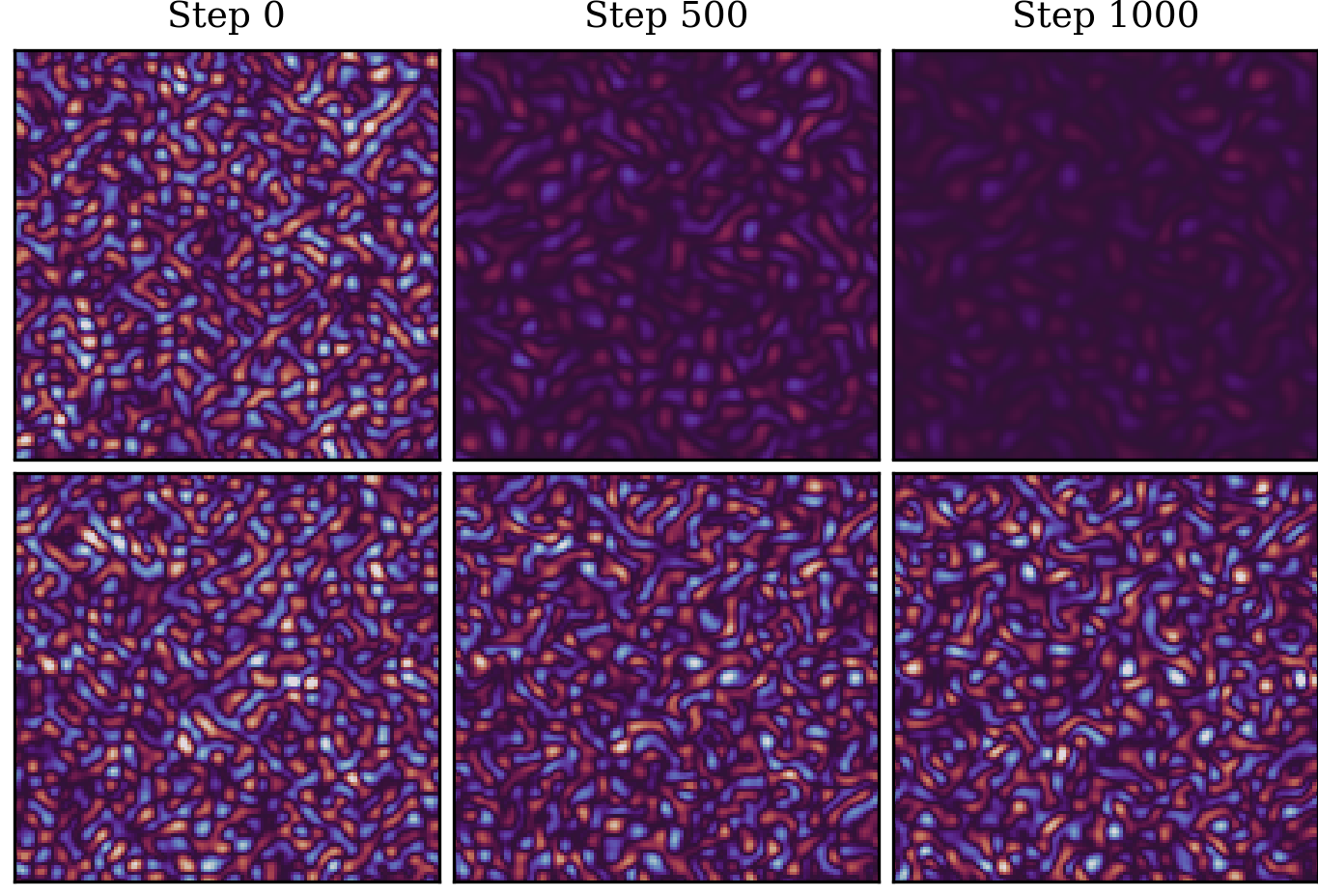}
     \end{minipage}

    \caption{\label{fig:forced}Energy spectrum $E[k]$ (left column) and evolution of vorticity fields in isotropic turbulence. Upper row: free simulation; lower row: restrained simulation.}
\end{figure}

Figure \ref{fig:forced} shows the vorticity fields and energy spectrum for a simulation at a resolution of 128$^2$ grid points with $Re=2000$ and $Ma=0.1.$ While the unrestrained simulation decays, the restrained simulation maintains the spectrum after an initial adjustment phase took place, starting from the artificial initialization field. This example shows that complicated forces are easily incorporated into simulations through automatic differentiation. This feature can be useful in many other applications.

\subsection{Benchmark }
\label{chapter:benchmark}
\begin{figure}[ht]
    \centering
    \includegraphics[width=0.99\linewidth]{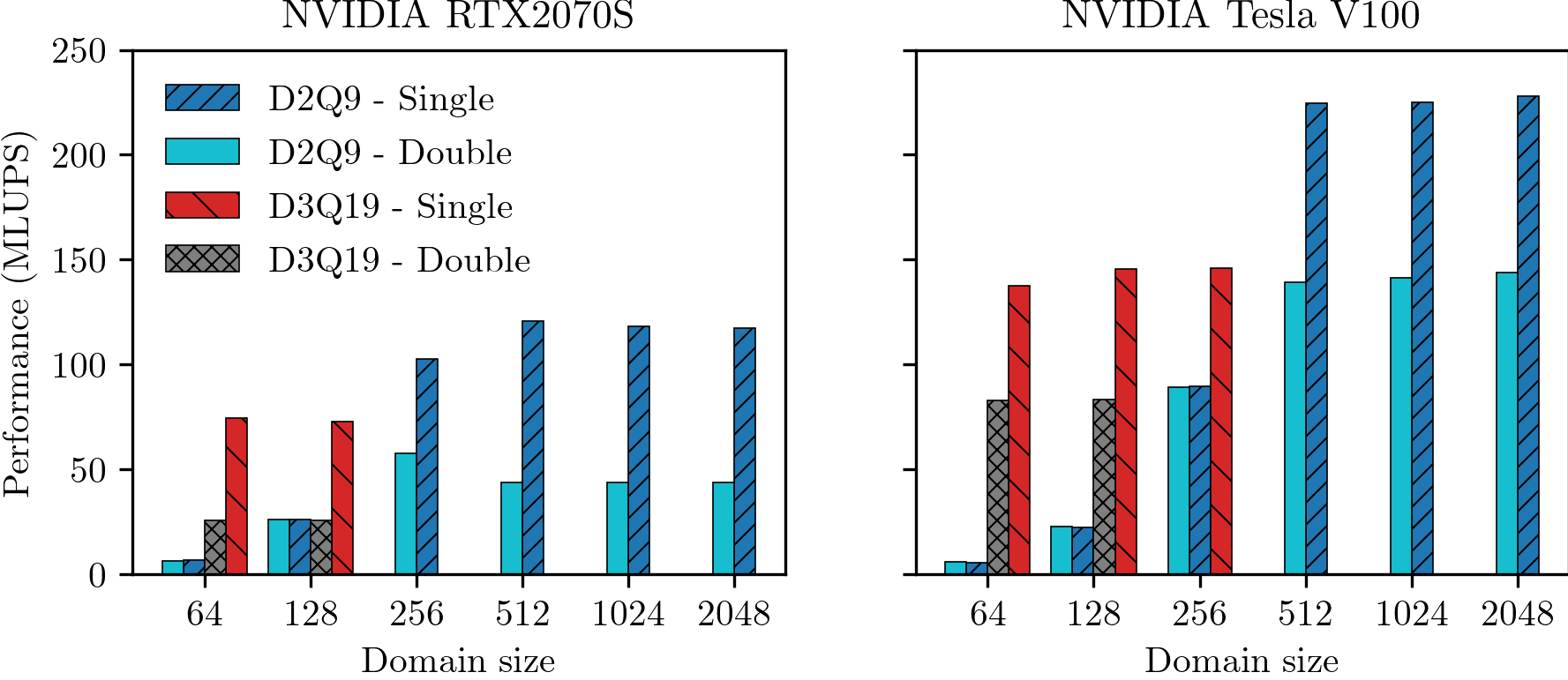}
    \caption{\label{fig:benchmark}Performance of Lettuce for simulating a Taylor-Green vortex.}
\end{figure}

\emph{Lettuce} attains a satisfactory performance due to the GPU operations provided by \emph{PyTorch}. The optimized backend code enables fast CUDA-driven simulations on both cluster GPUs and even standard GPUs for workstations. We evaluated the performance of \emph{Lettuce} by simulating a Taylor-Green vortex in both 2D (D2Q9) and 3D (D3Q19). The results are compared for both an NVIDIA Tesla V100 and an NVIDIA RTX2070S in single and double precision. Figure \ref{fig:benchmark} compares the performance in MLUPS (Million Lattice Updates Per Second) for different resolutions. 

With increasing domain size $64^2; 128^2; 256^2; 512^2$, the simulation speed increases to 140 MLUPS in two dimensions and 83 MLUPS in three dimensions using an NVIDIA Tesla V100 in double precision. The simulation speed increases by over $60 \si{\percent}$ to $75 \si{\percent}$ when calculations are performed in single precision. Using an off-the-shelve NVIDIA RTX2070S, the computational efficiency was lower, as expected. Performance peaked at 58 MLUPS using a domain size of $256^2$ in two dimensions. For higher resolutions, the performance decreased to 44 MLUPS. By using single precision, the performance can be increased by $180 \si{\percent}$ for higher resolutions. This comparison shows that even on a commercially available consumer-grade GPU high-performance simulations can be performed in an acceptable time. Further speedups will be obtained by implementing custom C++ and CUDA extensions, for which \emph{PyTorch} offers a modular and well-documented interface. Such extensions as well as a distributed multi-GPU implementation through \texttt{torch.distributed} are natural enhancements that are currently in progress.

\section{Conclusion}
\label{chapter:conclusion}
We have introduced \emph{Lettuce}, a \emph{PyTorch}-based lattice Boltzmann code that bridges lattice Boltzmann simulations and machine learning. We have demonstrated how simulations can be set up and run with minimal use of source code. This eases code development significantly and flattens the learning curve for beginners. Scientists and engineers can run GPU-accelerated three-dimensional simulations even on local workstations, which benefits rapid prototyping of lattice Boltzmann models. Besides machine learning routines, the framework supports automatic differentiation for flow control and optimization. As an example, a forced isotropic turbulence simulation was run with a maintained energy spectrum.  Furthermore, we have defined a neural collision model to demonstrate the benefits of incorporating neural networks into the lattice Boltzmann method. The presented results indicate that neural collision models can outperform traditional collision operators and reduce numerical errors, which motivates further research in this direction.

\bibliographystyle{splncs04}
\bibliography{literature}

\end{document}